\documentclass[preprint,showpacs,preprintnumbers,amsmath,amssymb,
nofootinbib,eqsecnum,12pt,floatfix]{revtex4}

\usepackage{graphicx}

\newlength{\dummysp}
\newcommand{\half}{\frac{1}{2}}

\newcommand{\beq}{\begin{eqnarray}}
\newcommand{\eeq}{\end{eqnarray}}

\newcommand{\s}{{\sigma}}
\newcommand{\psibar}{{\bar \psi}}
\newcommand{\Psibar}{{\bar \Psi}}
\newcommand{\chibar}{{\bar \chi}}
\newcommand{\vev}[1]{{\langle #1 \rangle}}

\newcommand{\ord}[1]{{{\cal O}(#1)}}
\newcommand{\gappeq}{\mathrel{\rlap {\raise.5ex\hbox{$>$}}
{\lower.5ex\hbox{$\sim$}}}}
\newcommand{\lappeq}{\mathrel{\rlap{\raise.5ex\hbox{$<$}}
{\lower.5ex\hbox{$\sim$}}}}
\newcommand{\myref}[1]{(\ref{#1})}

\newcommand{\ben}{\begin{enumerate}}
\newcommand{\een}{\end{enumerate}}
\newcommand{\bit}{\begin{itemize}}
\newcommand{\eit}{\end{itemize}}

\newcommand{\Rbf}{{\bf R}}

\newcommand{\Ncal}{{\cal N}}

\newcommand{\susy}{supersymmetry}

\newcommand{\lambar}{{\bar \lambda}}
\newcommand{\cond}{\vev{\lambar \lambda}}
\newcommand{\condt}{$\cond$}

\newcommand{\DMR}{DESY-M\"unster-Roma}
\newcommand{\mres}{m_{\text{res}}}

\newcommand{\mrest}{$\mres$}

\def\[{\left [}
\def\]{\right ]}
\def\({\left (}
\def\){\right )}

\begin{document}

\title{Lattice super-Yang-Mills \\ using
domain wall fermions in the chiral limit}

\author{Joel Giedt}
\email{giedtj@rpi.edu}
\affiliation{Department of Physics, Applied Physics and Astronomy,
Rensselaer Polytechnic Institute, 110 8th Street, Troy NY 12065 USA}

\author{Richard Brower}
\email{brower@bu.edu}
\affiliation{Physics Department, Boston University,
590 Commonwealth Avenue, Boston MA 02215}

\author{Simon Catterall}
\email{smc@physics.syr.edu}
\affiliation{Department of Physics, Syracuse University, 
Syracuse, NY 13244 USA}

\author{George T. Fleming}
\email{George.Fleming@yale.edu}
\affiliation{Department of Physics, Sloane Laboratory, 
Yale University, New Haven, Connecticut 06520, USA}

\author{Pavlos Vranas}
\email{vranasp@llnl.gov}
\affiliation{Physical Sciences Directorate,
Lawrence Livermore National Laboratory, 7000 East Ave., Livermore, CA 94550}

\date{Dec.~2, 2008}

\begin{abstract}
Lattice ${\cal N}=1$ super-Yang-Mills theory formulated 
using Ginsparg-Wilson fermions
provides a rigorous non-perturbative definition of the continuum theory
that requires no fine-tuning as the lattice spacing is reduced to zero. 
Domain wall fermions are one explicit scheme for achieving this
and using them we have performed large scale Monte Carlo
simulations of the theory for gauge group $SU(2)$.
We have measured the gaugino condensate,
static potential, Creutz ratios and residual mass
for several values of the domain wall separation $L_s$, 
four-dimensional lattice volume, and two values of the 
bare gauge coupling.
With this data we are able to extrapolate the gaugino
condensate to the chiral limit, to express it in physical
units, and to establish important benchmarks for future studies of 
super-Yang-Mills on the lattice.
\end{abstract}



\maketitle

\section{Introduction}
The lattice formulation of supersymmetric theories necessarily
breaks most or all\footnote{In cases with extended supersymmetry
it is sometimes possible to preserve a nilpotent subalgebra (see for example
the reviews \cite{Giedt:2006pd,Catterall:2005eh} and references therein).
For minimal supersymmetry, such as is studied here, all of the generators
are broken.} 
of the continuum supersymmetry, since the translation group,
a subgroup of the super-Poincar\'e group, is broken to a discrete subgroup.
This is intimately related to the failure of the Leibniz
rule on the lattice~\cite{Dondi:1976tx}.  
Hence all relevant and marginal operators
that are allowed by lattice symmetries will be generated
radiatively, so that generically the long distance effective theory will
have supersymmetry badly broken.

On the other hand, the only relevant or marginal operator allowed
in a gauge invariant lattice formulation of
pure $\Ncal=1$ super-Yang-Mills \cite{Ferrara:1974pu} (SYM) 
with hypercubic symmetry
is the gaugino mass term, as was emphasized long
ago in the analysis of \cite{Curci:1986sm}.  By using Ginsparg-Wilson 
fermions~\cite{Ginsparg:1981bj}, the associated
lattice chiral symmetry~\cite{Luscher:1998pqa} protects 
against additive renormalizations
of the gaugino mass in the continuum limit.  Hence the desired
continuum theory is obtained without fine-tuning of counterterms,
merely by setting the bare fermion mass to zero.

% HERE

A viable lattice formulation provides nonperturbative
information regarding SYM, the  foundation for all nonabelian four-dimensional supersymmetric
gauge theories.  Given the unique features of supersymmetric
theories, it makes sense to study their strong dynamics by
as many means as possible.  It is quite exciting that the
lattice approach can be brought to bear on $\Ncal=1$ SYM
in a rigorous and reliable way, since it would be enlightening
to study the nonperturbative aspects of this theory in
detail using numerical techniques.  For instance, information
on the spectrum and renormalization of nonholomorphic operators
would be welcome.

It is essential to quantify the size of lattice artifacts,
since these lead to deviations from the continuum physics.  One of the
purposes of the present study is to determine the regime
of lattice parameters where continuum SYM makes its appearance
(to a good approximation), and to characterize the difficulty 
of performing Monte Carlo
studies in that limit, given the algorithms and computing resources
that are currently available.  The study that we present is representative
of what can be achieved with a dedicated world-class supercomputing
resource---of order\footnote{This
is the {\it actual} computing power brought to bear.  The theoretical
computing power utilized was ten times this.} 
1 Teraflop/s $\times$ year (30 Million IBM BlueGene/L core hours)
on Rensselaer's Computational Center for Nanotechnology Innovation---and highly
optimized parallel code (a modified version of the Columbia
Physics System).  It will be seen that we are able to obtain
reliable extrapolations to the chiral limit (vanishing gaugino
mass), but that the study was quite demanding and could not
have been performed with anything less than the resources just described.

Whereas much is known about the vacuum structure
of SYM by continuum methods \cite{Veneziano:1982ah,Affleck:1983mk,
Affleck:1984xz,Novikov:1985ic,Davies:1999uw,Cachazo:2002ry},
nothing is known about its long distance
dynamics and spectrum.  Furthermore, lattice methods have the ability to
reveal far more detail about the vacuum, as one can subject it to
a configuration by configuration analysis, as has been done for pure
Yang-Mills with considerable success. 
Finally, we are interested in the effects of a small, nonzero
gaugino mass in the nonperturbative regime.  The explicit, but controllable,
chiral symmetry breaking of the fermion discretization we
use here allows us to explore the impact of
a small gaugino ``soft mass'' on nonperturbative quantities
such as the condensate, the string tension and the
shape of the static potential.  This is in the spirit of a large
number of proposals made over the last several years \cite{Evans:1997jy,
Farrar:1997fn,Farrar:1998rm,Gabadadze:1998bi,Cerdeno:2003us,
Merlatti:2004df,Auzzi:2005fi}.  Ultimately, the impact of nonzero
gaugino mass on the spectrum of bound states will emerge from high
statistics studies that are beyond the scope of the present
work.  It will then be possible to compare the lattice data
to the references just cited.

One implementation of Ginsparg-Wilson fermions is domain wall fermions (DWF) 
\cite{Kaplan:1992bt,Shamir:1993zy}
in the limit of infinite separation between the walls, $L_s \to \infty$.
Besides the absence of nonperturbative fine-tuning of the gaugino
mass, DWF have the advantage that the fermion
measure is real, positive and the square root of the determinant
which enforces the Majorana condition is analytic with a phase
that is independent of the gauge fields \cite{Neuberger:1997bg,Kaplan:1999jn}.
These three features are all lacking in the Wilson fermion formulation
that was applied in the only concerted effort to date to study SYM
on the lattice by the \DMR\ collaboration \cite{Campos:1999du,Farchioni:2001wx,
Farchioni:2004ej,Farchioni:2004fy,Pee03,DMR_other}
and to a lesser extent Donini et al.~\cite{Donini}. (Recently,
this program has been revived~\cite{DMR08}.)
Our research,
which has already appeared in preliminary form \cite{Gie08d}, is in some
sense a continuation of the work of
Fleming, Kogut and Vranas (FKV)~\cite{Fleming:2000fa} who pioneered the use of
DWF for studying ${\cal N}=1$ SYM. Similar work has been
initiated by Endres \cite{End08a}.
What sets the present study apart is that an extensive scan of the
domain wall separation $L_s$ and measurement of
the residual chiral symmetry breaking mass $\mres$ was done at two values of the
bare lattice gauge coupling ($\beta=4/g^2=2.3$ and $2.4$) 
and spatial/temporal volumes ($L^3=8^3$ and $16^3$; $T=16,32$).
This has allowed us to obtain a reliable
chiral extrapolation ($\mres \to 0$), and a preliminary view
on what occurs as we take the continuum, theormodynamic limit ($\beta,L,T \to \infty$).

The lattice formulation that is used in this study has already been
described by FKV~\cite{Fleming:2000fa};
it is reviewed in Appendix \ref{lf}.  In brief, the lattice employs 
Shamir DWF \cite{Shamir:1993zy} in the adjoint 
representation of SU(2), and the one-plaquette Wilson gauge action.
The Majorana condition is imposed through a square root on the fermion
determinant, which as mentioned above is analytic and
introduces no gauge field dependent sign ambiguity \cite{Neuberger:1997bg,Kaplan:1999jn}.  

All results reported in this article utilize a domain wall height 
$m_0$=1.9, as in the FKV simulations, though we will comment
briefly on some tests we did at other values of $m_0$.
Lattice configurations were generated with a
dynamical fermion mass $m_f=0$, so that the finite size of the fifth dimension,
parameterized by $L_s$,
was the sole infrared regulator, through the corresponding
additive mass correction $\mres$ (reviewed in Appendix \ref{smres}), which is a measure of
residual chiral symmetry breaking \cite{Blum:2000kn}.
One does not expect an optimal value of $m_0$ to exist \cite{Shamir:2000cf},
but for stronger couplings, the range is rather narrow.
Our $\mres$ measurements below show that we are in the correct
phase---the explicit chiral symmetry breaking decreases as
$L_s$ is increased---for $m_0=1.9$.  Finally, using the lattice configurations
that we generated, we computed $\mres$ for DWF propagators with
other values of $m_0$.  We found that $\mres$ could be lowered
slightly by increasing $m_0$ toward the critical value 2, and
that decreasing $m_0$ from 1.9 increased $\mres$.  The decrease
in $\mres$ by increasing $m_0$ was not significant, so we did
not pursue the issue further.

In Section \ref{bgco} we give our results for the bare gaugino
condensate for various couplings $\beta=4/g^2$ and domain
wall separations $L_s$.  Then in Section \ref{gluob} we summarize
our findings for ``gluonic'' observables (i.e., those related
to the nonabelian gauge bosons), principally
the string tension and thereby the Sommer parameter $r_0/a$.  
Next in Section \ref{extr}
we discuss the chiral extrapolations of the gaugino condensate based on our simulation
results.  We conclude in Section \ref{conc}.  In addition to the
two appendices mentioned above, Appendix \ref{ssim} describes
technical aspects of the simulation that may be of interest.

\section{Bare gaugino condensate}
\label{bgco}
We have validated our simulation code by comparing to FKV at several points.
We obtained results that agree with FKV, to within 1\% statistical errors.

A summary of all results obtained here for the gaugino
condensate \condt\ is given in Tables~\ref{consca}, 
\ref{tall2p3} and \ref{tall2p4}.  The residual chiral symmetry
breaking is parameterized through $\mres$~\cite{Blum:2000kn},
which we briefly review in Appendix \ref{smres}.
Measurements were conducted on large and small lattice volumes;
it can be seen that in lattice units the finite-size dependence is mild
or insignificant for $\beta=2.3$ but quite noticeable for $\beta=2.4$.
This is sensible, given that $\beta=2.4$ corresponds to a finer lattice spacing,
and hence the physical volumes are smaller.
Simulations on $16^3 \times 32$ volumes with $L_s=48$ 
are in progress and will be presented elsewhere.

\begin{figure}
\begin{center}
\includegraphics[width=2.5in,height=4in,angle=90]{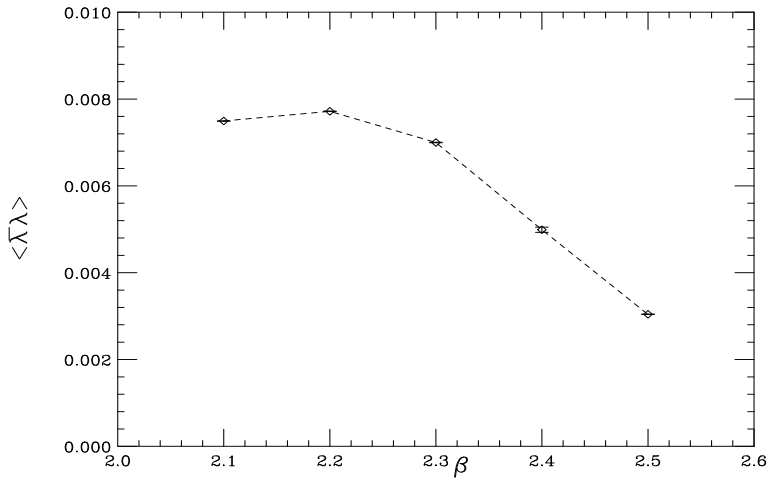}
\caption{Condensate vs.~$\beta$ for $16^3 \times 32$ lattice with
$L_s=16$.}
\label{cf1}
\end{center}
\end{figure}

For the $L_s=16$ lattices, which are relatively inexpensive, a scan over $\beta$
was performed, with results given in Table \ref{consca} and
shown in Fig.~\ref{cf1}.  The vanishing
extrapolated value at $\beta \sim 2.7$ is apparently due to finite-size effects
that cause the system to deconfine.

\begin{table}
\begin{center}
\begin{tabular}{|c|c|c|c|c|c|} \hline
$\beta$    & 2.1         & 2.2         & 2.3         & 2.4        & 2.5 \\ \hline
$\cond a^3$ & 0.007494(5) & 0.007719(5) & 0.007051(5) & 0.00499(6) & 0.003043(5) \\
\hline
\end{tabular}
\caption{A scan of the condensate versus $\beta$
for the $16^3 \times 32$ lattice with domain wall separation~$L_s=16$.
\label{consca}
}
\end{center}
\end{table}

\begin{table}
\begin{center}
\begin{tabular}{|c|c|c|c|c|c|c|} \hline
$V \times T$   & $L_s$ & $\mres a$ & $\cond a^3$ & $\mres r_0$ & $\cond r_0^3$  \\ \hline
$8^3 \times 8$   & 16 & 0.158(5) & 0.00711(7) & --- & ---  \\ % ok
$8^3 \times 32$  & 16 & 0.181(3) & 0.00703(4) & 0.75(13) & 0.51(27) \\
$16^3 \times 32$ & 16 & 0.184(2) & 0.007051(5) & 0.668(10) & 0.337(11)  \\ % 11/14/08
\hline
$8^3 \times 32$  & 24 & 0.1541(15) & 0.005112(8) & 0.610(97)  & 0.32(15)  \\ % 11/14/08
$16^3 \times 32$ & 24 & 0.1564(17) & 0.005321(9) & 0.546(55)  & 0.226(68) \\ % 12/1/08
\hline
$8^3 \times 32$ & 32 & 0.1319(12) & 0.004321(11) & 0.501(69) & 0.24(10)  \\ % 11/14/08
$16^3 \times 32$ & 32 & 0.143(2) & 0.00445(2) & 0.483(58) & 0.172(61)  \\ % 11/14/08
\hline
$8^3 \times 32$ & 40(I) & 0.1183(54) & 0.00383(3) & --- & ---  \\ \hline %
$8^3 \times 16$  & 48 & 0.1043(17) & 0.003563(20) & 0.361(31) & 0.148(37) \\ % 11/14/08
$8^3 \times 32$  & 48 & 0.1071(10) & 0.003551(11) & 0.409(31) & 0.198(45)  \\ \hline % 11/14/08
$8^3 \times 32$  & 64 & 0.08864(84) & 0.003164(10) & 0.300(35) & 0.122(42) \\ % 11/14/08
\hline
\end{tabular}
\caption{The gaugino condensate \condt\ and residual mass $\mres$ for various lattice 
sizes and $L_s$ values, all at $\beta=2.3$.
The $L_s=40$ value, with an ``(I)'' after it, is obtained by the
interpolation method described in the text,
and summarized in Table~\ref{sqc}.  
Values in units of the Sommer parameter $r_0$ are also shown, for those cases where the
potential was measured (in particular, for all points that are included in the
chiral extrapolation fit).  The $L_s=16$ data was not included
in the linear chiral extrapolation fit, because these points had
too much curvature (with respect to $\mres a$) associated with them.
\label{tall2p3} }
\end{center}
\end{table}

\begin{table}
\begin{center}
\begin{tabular}{|c|c|c|c|c|c|} \hline
$V \times T$ & $L_s$ & $\mres a$ & $\cond a^3$ & $\mres r_0$ & $\cond r_0^3$ \\ \hline
$8^3 \times 32$  & 16 & 0.080(2)  & 0.004839(15) & 0.547(30) & 1.55(22) \\
$16^3 \times 32$ & 16 & 0.0969(8) & 0.00499(6)   & 0.5355(66) & 0.842(25) \\
\hline
$8^3 \times 32$  & 24 & 0.0601(15) & 0.003293(17) & 0.417(26) & 1.10(18) \\ % 12/2/08 
$16^3 \times 32$ & 24 & 0.0838(17) & 0.00389(8) & 0.385(35) & 0.38(10) \\ % 11/14/08
\hline
$16^3 \times 32$ & 28(I) & 0.0721(33)  & 0.003452(45)  & --- & ---  \\ %
\hline
$8^3 \times 32$  & 32 & 0.0486(12) & 0.00269(2) & 0.296(15) & 0.61(08)  \\ % 11/14/08
$16^3 \times 32$ & 32 & 0.0653(15) & 0.003330(12) & 0.313(33) & 0.37(11) \\ % 11/14/08
\hline
$8^3 \times 32$  & 40(I) & 0.0390(24) & 0.00234(8) & --- & --- \\ % 
\hline
$8^3 \times 32$  & 48 & 0.0328(9) & 0.002165(18) & 0.224(17) & 0.69(15) \\  % 11/14/08
\hline
\end{tabular}
\caption{Results similar to Table \ref{tall2p3},
except that these are for $\beta=2.4$.
\label{tall2p4} }
\end{center}
\end{table}

We have measured the condensate at other values of $L_s$ using
a sea-$L_s$/valence-$L_s$ approach.  The condensate was
measured using DWF with $L_s^{\text{val.}}$ on top of dynamical lattices produced
using a nearby $L_s^{\text{sea}}$.  Performing this for $L_s^{\text{sea}}$
values on either side of $L_s^{\text{val.}}$ yields robust interpolated (I)
results, as can be seen in Table \ref{sqc}.  These are then
used in our fits of $\cond$ vs.~$\mres$, together with the
strictly dynamical ($L_s^{\text{val.}}=L_s^{\text{sea}}$)
measurements of Tables \ref{tall2p3} and \ref{tall2p4}.
Interpolations (``I'') are taken halfway between the results. Half
the difference plus the statistical errors added in quadrature
is used as an error estimate for the interpolation.

\begin{table}
\begin{center}
\begin{tabular}{|c|c|c|c|c|c|c|} \hline
$\beta$ & $V$ & $T$ & $L_s^{\text{val.}}$ & $L_s^{\text{sea.}}$ & $\mres a$ & $\cond a^3$ \\ \hline
2.3 & $8^3$  & 32 & 40 & 32   & 0.117(4) & 0.003818(9) \\   % 
2.3 & $8^3$  & 32 & 40 & 48   & 0.1196(10) & 0.003843(9) \\ % 
2.3 & $8^3$  & 32 & 40 & I    & 0.1183(54) & 0.00383(3) \\ \hline
2.4 & $16^3$ & 32  & 28 & 24 & 0.0707(13) & 0.003407(3) \\ % 
2.4 & $16^3$ & 32  & 28 & 32 & 0.0734(15) & 0.003496(3) \\ %
2.4 & $16^3$ & 32  & 28 & I  & 0.0721(33) & 0.003452(45)  \\ % 
\hline
2.4 & $8^3$  & 32  & 40 & 32 & 0.0381(10) & 0.002284(13) \\
2.4 & $8^3$  & 32  & 40 & 48 & 0.0398(11) & 0.002397(17) \\
2.4 & $8^3$  & 32  & 40 & I  & 0.0390(24) & 0.00234(8) \\
\hline
\end{tabular}
\caption{The valence/sea results and interpolations (``I'').  
Here $L_s^{\text{val.}}$ is the value used for the measurements
and $L_s^{\text{sea}}$ is the value used in the dynamical fermion
simulations.
\label{sqc}}
\end{center}
\end{table}

We also use the results of Section \ref{stpo} below to express
$\mres$ and $\cond$ in terms of the Sommer scale $r_0$ \cite{Sommer:1993ce}.
Note that the $\beta=2.4$ value of $\mres r_0$ at $L_s=48$
indicates that the effective gaugino mass (which should
be approximately equal to $\mres$) is roughly 1/4 the inverse
Sommer scale, so that we are beginning to enter
the chiral regime where \susy\ is well approximated.
On the other hand, it can be seen that $\mres r_0$ is
unpleasantly large for $\beta=2.3$ with $L_s \leq 32$, and likewise
the condensate in physical units is small compared to
the $\beta=2.4$ results.  Clearly $\beta=2.3$ is further away from the 
supersymmetric limit due to the coarser lattice.  On the other hand
it can be seen that the $\beta=2.4$ data shows a marked
volume dependence due to the smaller physical ``box'' that
the states must squeeze into.

We note that in the present context the $\mres$ measurement coming from the midpoint
``pion'' propagator calculation is a quenched probe of
explicit chiral symmetry breaking,
since in the dynamical theory (i.e., the one pertaining
to the lattice action that is used to generate configurations)
we do not have two ``valence quarks'' and a nonanomalous
continuous chiral symmetry that would give rise to pseudo-Nambu-Goldstone
bosons.  Rather, the (adjoint) pions that are measured in the
midpoint calculation of $\mres$ are pseudo-Nambu-Goldstone
bosons of an $SU(4|3)$ graded chiral symmetry that is spontaneously
broken in the chiral limit, and explicitly broken
at finite $L_s$, as we now explain.

The $SU(4)$ subalgebra of the graded Lie algebra $SU(4|3)$ is associated
with the gaugino (a Majorana fermion) plus three quenched Majorana
fermion probes.  The total of four Majorana fermions with
degenerate mass is equivalent to two Dirac fermions in 
the adjoint representation, with a resulting
$SU(4)$ chiral flavor symmetry in the chiral limit.   The fact that three of the
four Majorana degrees of freedom are quenched is equivalent to introducing
three Majorana ghosts, also with the same mass.  As in
partially quenched QCD, the ghosts cancel the contribution
of the nondynamical fermions to the functional integration
measure.  Also analogous to quenched QCD, there is a graded Lie
algebra that relates the fermion and ghost fields, namely
$SU(4|3)$ in the present case.  The PCAC mass ($\mres$) associated 
with this $N_f=2$ adjoint-Dirac fermion chiral symmetry 
breaking is a good probe of the DWF 
chiral limit, for the same reasons that it is a solid
tool in quenched DWF-QCD studies.  Investigations
of the effective theory description of the $SU(4|3)$
algebra, from the theoretical perspective as it relates to DWF, are 
in progress \cite{Gie09a}.  Finally we note that in
recent spectrum studies of one-flavor QCD, a similar
non-singlet flavor current was utilized with success \cite{Farchioni:2008na}.

\section{Gluonic observables}
\label{gluob}
One of the interesting features of SYM is that it is
a theory with dynamical fermions that do not screen
static sources in the fundamental representation.
In contrast to QCD, it has true confinement in the
sense of an area law and no string breaking. (Recall that the
gauge action is expressed in terms of fundamental links, so we are not
studying the $SU(2)/Z_2=SO(3)$ gauge theory, wherein fundamental
sources would have an ambiguous meaning.) On the
lattice, we can therefore study a very interesting
static potential---one with chiral fermions and a
nonvanishing string tension.  These features of SYM
will be presented here, illustrating the special ability
of the lattice approach:  to conduct detailed studies
of the nonperturbative aspects of the theory that the  continuum
methods cannot touch upon.

\subsection{Creutz ratios}
\label{strt}
\begin{figure}
\begin{center}
\includegraphics[width=3in,height=5in,angle=90]{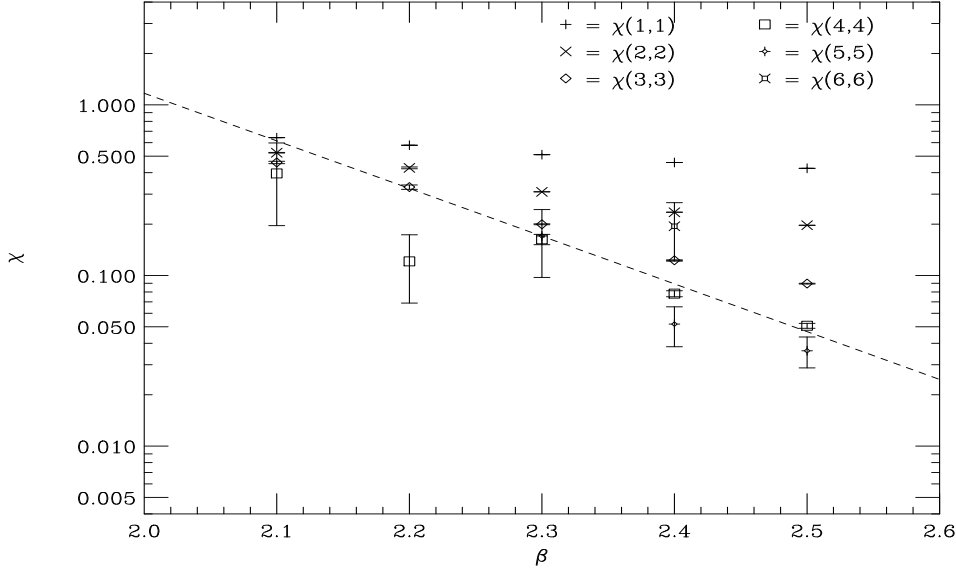}
\caption{Creutz ratios for the $16^3 \times 32 \times 16$ lattice.
The dashed line indicates the 2-loop prediction for
the dependence $a^2(\beta)$, apart from an overall normalization
that is determined by requiring that the curve pass through the data.
\label{cf16}
}
\end{center}
\end{figure}

Here we look at Creutz ratios \cite{Creutz:1980wj} as a probe of the string
tension in lattice units, $\s a^2$, as well as to delineate the scaling
regime where the continuum limit may be extracted.
Results for the  $16^3 \times 32 \times 16$ lattice are shown
in Table \ref{ct16} and Fig.~\ref{cf16}.  Although the errors
are somewhat large, scaling is clearly setting in at around $\beta\sim 2.4$
as can be seen by the $\chi(4,4)$ ratios, which lie quite close to
the 2-loop curve.  Much beyond that $\beta$, finite size effects will
take over and it is necessary to move to a larger lattice.
For this reason, most of our simulations have been performed
at $\beta=2.3$ and $\beta=2.4$.

\begin{table}
\begin{center}
\begin{tabular}{|l|l|l|l|l|l|l|} \hline
$\beta$ & $\chi(1,1)$ & $\chi(2,2)$ & $\chi(3,3)$ 
& $\chi(4,4)$ & $\chi(5,5)$ & $\chi(6,6)$\\ \hline
2.1 & 0.6423(6)   & 0.5242(16) & 0.459(7)   & 0.4(2)    & --- & --- \\ \hline
2.2 & 0.580(2)    & 0.427(5)   & 0.330(10)  & 0.12(5)   & --- & --- \\ \hline
2.3 & 0.51051(11) & 0.3091(4)  & 0.1997(12) & 0.163(11) & 0.17(7) & --- \\ \hline
2.4 & 0.45966(7)  & 0.2346(3)  & 0.1227(9)  & 0.078(3)  & 0.052(14) & 0.19(7) \\ \hline
2.5 & 0.42493(8)  & 0.1969(3)  & 0.0896(5)  & 0.0507(17) & 0.036(7) & --- \\ \hline
\end{tabular}
\caption{Creutz ratios for the $16^3 \times 32 \times 16$ lattice. \label{ct16}}
\end{center}
\end{table}

\subsection{Static potential}
\label{stpo}
The static potential was obtained by measuring Wilson
loops with one side of length $t$ in the temporal direction, according
to standard methods.  Having obtained $V(r) a$ from fitting
the exponential decay in time, we next analyze the potential
in terms of the distance $ra$.  We fit the data to the standard form
\beq
V(r) a = V_0 a + \s a^2 (r/a) - \frac{\alpha}{r/a},
\eeq
as was also done by the \DMR\ collaboration in \cite{Campos:1999du},
and in the recent work~\cite{End08a}.
We obtain the Sommer parameter $r_0/a$ from this fit, using the formula
\beq
\frac{r_0}{a} = \sqrt{\frac{1.65-\alpha}{\s a^2}}.
\eeq
This approach to the determination of $r_0/a$ has
some sensitivity to the range of radii that is fit,
and obviously depends on what form we assume for $V(r) a$.
The force method of Sommer~\cite{Sommer:1993ce} 
improves on this by avoiding a static potential fit,
relying instead on tree-level improved finite differences.
Preliminary studies that we have conducted indicate
that smearing of gauge links and higher statistics will be
indisposable to a successful application of the
method, and so we will perform this in a future study.

The results of our static potential fit and
the derived quantities are presented in Tables~\ref{stpr_2p3} and \ref{stpr_2p4}.
Note that the values of the string tension in
physical units, $\s r_0^2$, are quite uniform for fixed $\beta$,
and have small errors.
For the $L=16$ results,
the fits were also done using the same set of Wilson loops as
in the $L=8$ case, denoted ``$L=8$ method'', so that dependence
choice of Wilson loops could be controlled
for, and therefore ruled out as a spurious source of finite
size dependence.  It also gives a clearer picture of the systematic
uncertainties associated with the measurement and fitting methods
for the static potential.

These results are to be compared with Table 1 of \cite{End08a}.
There, a nonzero fermion mass $m_f=0.02$ was used, $L_s=16$,
and somewhat different fit ranges for $r,t$ were employed.
In particular, our fits include the
points $r/a=1,\sqrt{2}$, which have very small errors and
thus strongly influence the fit.  Thus for $L_s=16$
we have also performed a fit with $r_{\min} = \sqrt{3}$ as 
was done in \cite{End08a}, as can be seen in the second $L_s=16$
entry for $\beta=2.3,2.4$ in Tables \ref{stpr_2p3} and \ref{stpr_2p4}.  
Our results for the fit quantities at $\beta=2.3$ are
in good agreement once this restriction is imposed.
For our other $L_s$ values we have far fewer samples,
as larger $L_s$ comes at greater computing cost.  The degradation
of statistical errors that results if we exclude the
$r=1,\sqrt{2}$ points is unacceptable, which is why we do
not quote results with the same $r_{\min} = \sqrt{3}$ as \cite{End08a}
for the other $L_s$ values.  On the other hand, it can be
seen from the $L_s=16$ results that the choice of $r_{\min}$
only has a 10\% effect on the $r_0/a$ estimate, so that the
choice of $r_{\min}$ is not crucial to the broad picture that
we are after in this preliminary work.
The $\beta=2.4$ results are also in reasonable
agreement with \cite{End08a}, comparing to the numbers
we obtain at $16^3 \times 32$, $L_s=16$, with $r_{\min}=\sqrt{3}$,
and keeping in mind the nonzero $m_f$ in \cite{End08a}.

\begin{table}
\begin{center}
\begin{tabular}{|c|c|c|c|c|c|c|c|}  \hline
$V \times T$ & $L_s$ & $V_0 a$ & $\s a^2$ & $\alpha$ & $r_0/a$  & $\s r_0^2$ & method \\ 
\hline
$8^3 \times 32$ & 16 & 0.717(83) & 0.074(28) & 0.368(55) & 4.16(73) & 1.282(55) & $L=8$ \\
$16^3 \times 32$ & 16 & 0.6533(80) & 0.1004(25) & 0.3271(57) & 3.630(39) & 1.3229(57) & $L=16$ \\
$16^3 \times 32$ & 16 & 0.489(45) & 0.1367(88) & 0.159(55) & 3.303(50) & 1.491(55) & $L=16,r\geq\sqrt{3}$ \\ 
\hline
$8^3 \times 32$  & 24 & 0.718(89) & 0.082(30) & 0.371(60) & 3.96(63) & 1.279(60) & $L=8$ \\
$16^3 \times 32$ & 24 & 0.752(70) & 0.102(24) & 0.411(46) & 3.49(35) & 1.239(46) & $L=16$ \\
$16^3 \times 32$ & 24 & 0.696(67) & 0.119(22) & 0.372(46) & 3.27(24) & 1.278(46) & $L=8$ \\
\hline
$8^3 \times 32$  & 32 & 0.748(82) & 0.087(27) & 0.400(55) & 3.80(52) & 1.250(55) & $L=8$ \\
$16^3 \times 32$ & 32 & 0.745(90) & 0.109(31) & 0.412(59) & 3.38(40) & 1.238(59) & $L=16$ \\
$16^3 \times 32$ & 32 & 0.635(50) & 0.146(17) & 0.338(33) & 3.00(14) & 1.312(33) & $L=8$ \\
\hline
$8^3 \times 16$  & 48 & 0.706(68) & 0.107(22) & 0.372(47) & 3.46(29) & 1.278(47) & $L=8$ \\
$8^3 \times 32$  & 48 & 0.768(47) & 0.085(15) & 0.414(33) & 3.82(29) & 1.236(33) & $L=8$ \\
\hline
$8^3 \times 32$  & 64 & 0.680(94) & 0.113(32) & 0.353(63) & 3.38(39) & 1.297(63) & $L=8$ \\
\hline
\end{tabular}
\caption{Gluonic observables obtained from the static potential for $\beta=2.3$.
We note that there is no sign of volume dependence in the string tension 
results expressed in physical units $\s r_0^2$.  Errors estimates
are obtained from a jackknife analysis of fits.  
\label{stpr_2p3}
}
\end{center}
\end{table}

\begin{table}
\begin{center}
\begin{tabular}{|c|c|c|c|c|c|c|c|}  \hline
$V \times T$ & $L_s$ & $V_0 a$ & $\s a^2$ & $\alpha$ & $r_0/a$  & $\s r_0^2$ & method \\ \hline
$8^3 \times 32$ & 16 & 0.617(11) & 0.0292(30) & 0.2857(83) & 6.84(33) & 1.3643(83) & $L=8$ \\
$16^3 \times 32$ & 16 & 0.5846(32) & 0.04531(91) & 0.2659(24) & 5.526(51) & 1.3841(24) & $L=16$ \\
$16^3 \times 32$ & 16 & 0.537(11)  & 0.0554(20)  & 0.219(15)  & 5.083(67) & 1.431(15) & $L=16, r \geq \sqrt{3}$ \\
\hline
$8^3 \times 32$ & 24 & 0.636(12) & 0.0280(33) & 0.2997(88) & 6.94(39) & 1.3503(88) & $L=8$ \\
$16^3 \times 32$ & 24 & 0.579(40) & 0.065(13) & 0.272(27) & 4.60(41) & 1.378(27) & $L=16$ \\
\hline
$8^3 \times 32$  & 32 & 0.609(12) & 0.0369(36) & 0.2809(90) & 6.09(28) & 1.369(90) & $L=8$ \\
$16^3 \times 32$ & 32 & 0.611(43) & 0.059(13)  & 0.295(29)  & 4.79(50) & 1.355(29) & $L=16$ \\
\hline
$8^3 \times 32$  & 48 & 0.648(15) & 0.0288(44) & 0.309(11) & 6.83(49) & 1.341(11) & $L=8$ \\
\hline
\end{tabular}
\caption{Gluonic observables obtained from the static potential for $\beta=2.4$.
We note that $r_0/a$ shows significant volume dependence for $L_s=32$.  
\label{stpr_2p4}
}
\end{center}
\end{table}

Above, we have used the results of Tables \ref{stpr_2p3} and \ref{stpr_2p4}
to scale the residual mass and condensate
to $r_0$ units.  (Note that the $r_0/a$ values with identical
lattice parameters were used in this procedure,
rather than a uniform $r_0/a$ value across all $\mres$ and $\cond$.)
With the string tension in hand, we now see that the energy
scale of confinement $\sqrt{\s r_0^2} \approx 1.4$ lies above the explicit chiral symmetry
breaking scale $\mres r_0$ by a factor of 1.8 to 3.8 for $\beta=2.3$, and 3.0 to 5.2
for $\beta=2.4$.  This is consistent with the observation that the
string tension results in Tables \ref{stpr_2p3} 
and \ref{stpr_2p4} are insensitive to the range of
$L_s$ values displayed there, when expressed in physical units ($\s r_0^2$).
That is, confinement dynamics are to
a good approximation decoupled
from the explicit chiral symmetry breaking.  Since the lowest lying excitations
of SYM are glueballs and superpartners, the gap associated with confinement
should also decouple these states from the explicit chiral symmetry
breaking.  Thus it appears that we are well into the regime where the spectrum
reflects supersymmetry, and it will be quite interesting to examine the
spectrum in order to check whether or not this is true---something
we will do in future work.  In addition,
it gives us confidence that we are performing the chiral extrapolation
of the condensate (next section) correctly, where the data points are dominated
by the physics of the supersymmetric theory.

\section{Extrapolation of the gaugino condensate}
\label{extr}
\begin{figure}
\begin{center}
\includegraphics[width=3in,height=5in,angle=90]{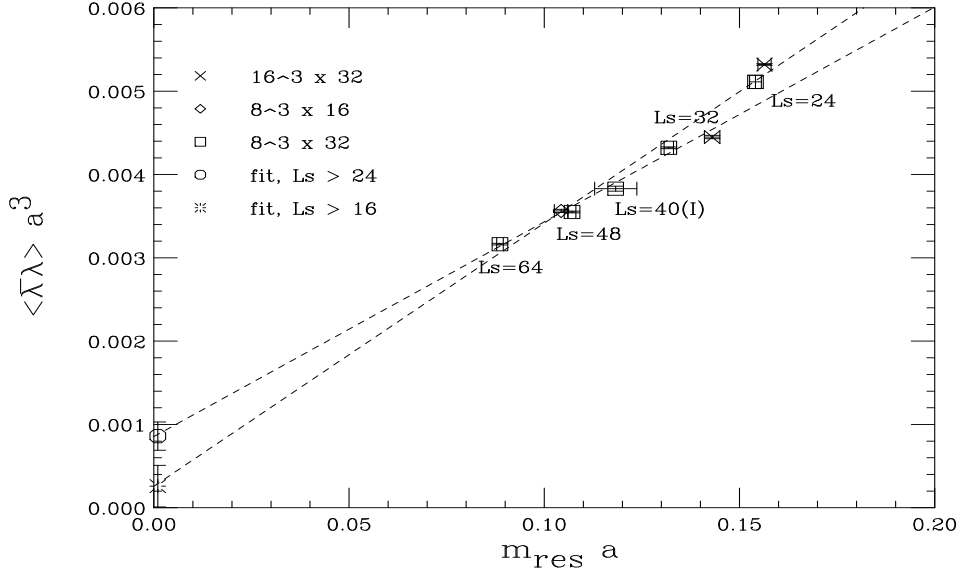}
\caption{Condensate vs.~\mrest\ for $\beta=2.3$, in bare lattice
units.  Dashed lines show the two linear fits (differing
by the minimum $L_s$ included).  Extrapolated values
together with fit errors are shown at $\mres=0$. \label{cvm2p3} }
\end{center}
\end{figure}

\begin{figure}
\begin{center}
\includegraphics[width=3in,height=5in,angle=90]{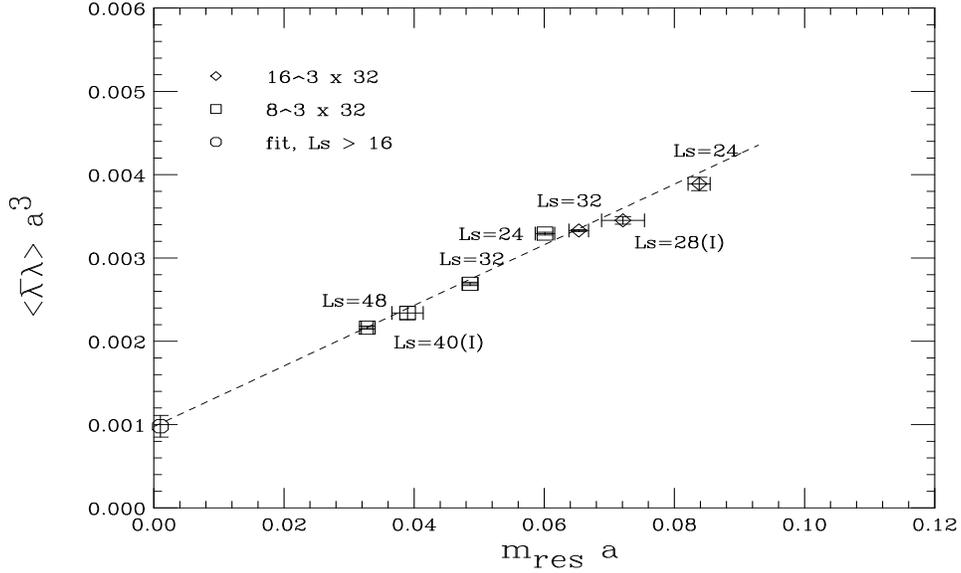}
\caption{Condensate vs.~\mrest\ for $\beta=2.4$, in bare lattice
units.  The dashed line shows the linear fit.  The extrapolated
value together with fit error is shown at $\mres=0$.  \label{cvm2p4} }
\end{center}
\end{figure}

One important question is the size of $L_s$ necessary to get into the linear
regime where 
\beq
\cond \approx c_0 + c_1 \mres
\label{conex}
\eeq
is a good approximation.  Obviously, this serves as an indicator of
where we need to be in order to have SYM well-approximated.  Thus,
the measurement of \condt\ vs.~\mrest\ is an important benchmark
for determining the regime in which other SYM phenomena can be studied
with the DWF lattice approach.
Another question is the extent to which $c_{0,1}$ are sensitive to
finite spacetime volume ($V_4=V \times T$ in our notation).  In fact,
we find, interestingly, that most of the volume dependence is absorbed into \mrest.
All of this is clearly seen from Figs.~\ref{cvm2p3} and \ref{cvm2p4}.  One sees that
to a good approximation, the $8^3 \times 32$ and $16^3 \times 32$
lattice data lie on the same line.  The smaller value of \mrest\ on
the smaller lattice is most likely due to a smaller density of
near-zero modes.  The chiral extrapolation ($\mres \to 0$) of $\cond a^3$
obtained from the fit is given in Table \ref{tbex}.  A feel
for the sensitivity to the fitted range of $L_s$ can be
seen from the two results we provide for $\beta=2.3$,
which differ by the minimum $L_s$ that was included.  
In fact, the quality of the $L_s>16$ fit is very poor
due to nonlinear dependence on $\mres$ that enters at $L_s=24$,
as can also be seen from Fig.~\ref{cvm2p3}.

\begin{table}
\begin{center}
\begin{tabular}{|c|c|c|} \hline
$\beta$ & $L_s > 24$ & $L_s > 16$ \\ \hline
2.3 & 0.00086(17) & 0.00026(25) \\ % 12/3/08
2.4 & --- & 0.00098(13) \\ % 12/3/08
\hline
\end{tabular}
\caption{Fit results for the chiral extrapolation
of the gaugino condensate, depending upon the range of $L_s$ values used.
For $\beta=2.3$, the quality of the $L_s>16$ fit is very poor
due to nonlinear dependence on $\mres$ that enters at $L_s=24$,
as can also be seen from Fig.~\ref{cvm2p3}.
\label{tbex}}
\end{center}
\end{table}

According to Table \ref{stpr_2p4}, the value of the lattice spacing $a$ is smaller
on the $\beta=2.4$ lattice.
Thus it is surprising that the extrapolated value of $\cond a^3$ is
larger for $\beta=2.4$ than for $\beta=2.3$.  On the other hand,
we note from Table \ref{tall2p4} that the $\beta=2.4$ condensates measured in physical
units ($\cond r_0^3$) are significantly 
larger than the $\beta=2.3$ ones, given in Table \ref{tall2p3}.  
A plausible interpretion is that there are larger renormalizations 
of the condensate on a finer lattice ($\beta=2.4$),
a hypothesis that we are preparing to test with
nonperturbative renormalization \cite{Martinelli:1994ty,Blum:2001sr} in an upcoming study.  
Under this interpretation
we have an understanding of why in Table \ref{tbex} the chirally extrapolated value of
the lattice units condensate for $\beta=2.4$ is larger
than the one for $\beta=2.3$, which does not agree
with naive expectations.  It can also be seen, for instance from the $\beta=2.4$, $L_s=32$ data
in Table \ref{tall2p4}, that a very significant finite size effect occurs for the condensate
expressed in physical units, a reflection of the finite
size dependence of the static potential fit results for this choice
of parameters (cf.~Table \ref{stpr_2p4}).
Finally, it is amusing that the extrapolated value of the $\beta=2.3$, $L_s>24$ fit
is within 1$\sigma$ of the extrapolated value of the $\beta=2.4$ fit,
though there is no reason why this should be true.

\section{Conclusions}
\label{conc}
We have performed a detailed Monte Carlo simulation
study of the supersymmetric limit of $\Ncal=1$, SU(2)
super-Yang-Mills using a DWF lattice formulation.
Our work follows from an earlier calculation by FKV
but significantly extends that work; 
we use larger lattices with two lattice spacings and are able to
probe much closer to the chiral limit.
Our results for the gaugino condensate show
the correct theoretical dependence on the residual mass and allow for
a reliable extrapolation to the chiral limit. Our results provide strong
evidence for a nonzero gaugino condensate in the supersymmetric continuum
limit, and establish important  benchmarks for future studies.

Future work that is envisioned is aimed at
developing a deeper understanding of the configurations
that are responsible for generating the nonzero gaugino condensate.
In particular, we would like to elucidate the continuum
picture on the cylinder $\Rbf^3 \times S^1$,
where it is monopoles and ``KK monopoles'' that
combine to yield the infinite volume value \cite{Davies:1999uw}.

At the same time, two important studies need to be done in order
to further develop the lattice results presented here.  First, we will
make a more accurate determination of the Sommer scale $r_0$, using
the force technique.  In that study, smearing of the gauge links
and other refinements appearing in \cite{Sommer:1993ce} will be used.
Second, nonperturbative renormalization of the gaugino condensate
will be performed \cite{Martinelli:1994ty,Blum:2001sr}.

Finally, we  note that rather large $L_s$ values were required
in order to get $\mres r_0 \sim 1/4$.  To improve the situation
we envision switching to simulations with modified versions of DWF
that have superior chiral behavior \cite{Vranas:2006zk,Brower:2004xi}.

\section*{Acknowledgements}
The research described here was made possible by the
supercomputing resource that was provided
by Rensselaer:  ten months of steady access to between
one and four IBM BlueGene/L racks at the Computational
Center for Nanotechnology Innovation (CCNI), for
a total of approximately 30 Million IBM BlueGene/L core hours.  This center was
built in a three-way partnership between Rensselaer, IBM and the State of New York.
Our project benefitted from reference to a copy of the 1999-2000
modified Columbia Physics System (CPS)
code used by Fleming, Kogut and Vranas \cite{Fleming:2000fa},
and was built by modification of the publically available 2008
CPS lattice QCD code, developed in part under a Department of Energy
SciDAC grant.
J.G.~thanks Adam Todorski at CCNI, and Chulwoo Jung
at Columbia University and Brookhaven National Laboratory
for practical help on many occasions.  He also
received assistance from faculty development funds at Rensselaer.
RCB's research is supported in part by the Department of Energy under
contract No.~DE-FG02-91ER40676.  SMC is supported in part by the
Department of Energy under contract No.~DE-FG02-85ER40237.
Finally, we thank Michael Endres for correcting an error
in the first version of this article.

\appendix

\section{Lattice formulation}
\label{lf}

In this appendix, the ${\cal N}=1$, SU(2) SYM lattice action and
operators associated with the gaugino condensate are
described.  The DWF formulation for this
theory is identical to \cite{Neuberger:1997bg,Kaplan:1999jn,Fleming:2000fa}, 
and is described here for completeness.
The lattice consists of an SU(2) gauge theory with a single Majorana
fermion in the adjoint representation.  As such, the fermionic part of the path
integral is the analytic square root of the
Dirac determinant, yielding the Pfaffian of the corresponding matrix.

The partition function is
\begin{equation}
Z = \int [dU] \int [d\Psi] \int [d\Phi] e^{-S} .
\label{Z}
\end{equation}
Here $U_\mu(x)$, $\mu=1,2,3,4$ is the four-dimensional gauge link field in the
fundamental representation, $\Psi(x,s)$ is a (real) five-dimensional
Majorana spinor in the adjoint representation and $\Phi(x,s)$ is a
(real) five-dimensional bosonic Pauli Villars (PV) field with the
same indices as the Majorana fermion.  The coordinate $x$ denotes sites in the
four-dimensional spacetime box, which has $L$ sites in spatial
directions and $T$ sites in time.  The boundary conditions along these directions are taken
to be periodic for the gauge link fields but antiperiodic
in time for the fermion and Pauli-Villars fields.  
The coordinate of the fifth direction
is $s=0,1, \dots, L_s-1$, where $L_s$ is the size of that direction
and is taken to be an even number.  The action $S$ is given by:
\begin{equation}
S = S_G(U) + S_F(\Psi, U) +
S_{PV}(\Phi, U) \ .
\label{action}
\end{equation}

$S_G(U)$ is the pure gauge part and is defined
using the standard single plaquette action of Wilson:
\begin{equation}
S_G = \beta \sum_p ( 1 - {1 \over 2} {\rm Re Tr}[U_p])
\label{action_G}
\end{equation}
where $\beta = 4/g^2$ and $g$ is the gauge coupling.

The fermion part $S_F(\Psi, U)$ is given by:
\begin{equation}
S_F = - \sum_{x,x^\prime,s,s^\prime} \Psibar(x,s) D_F(x,s; x^\prime,
s^\prime) \Psi(x^\prime,s^\prime)
\label{action_F}
\end{equation}
where $D_F$ is the DWF Dirac operator in the form of \cite{Furman:1994ky}:
\begin{equation}
D_F(x,s; x^\prime, s^\prime) = \delta(s-s^\prime) {\not \hspace{-3pt} D} (x,x^\prime)
+ {\not \hspace{-3pt} D}^\bot(s,s^\prime) \delta(x-x^\prime)
\label{D_F}
\end{equation}
\begin{eqnarray}
{\not \hspace{-3pt} D} (x,x^\prime) &=& {1\over 2} \sum_{\mu=1}^4 \left[ (1+\gamma_\mu)
V_\mu(x) \delta(x+\hat\mu - x^\prime) + (1-\gamma_\mu)
V^\dagger_\mu(x^\prime) \delta(x^\prime+\hat\mu - x) \right] \nonumber \\
&+& (m_0 - 4)\delta(x-x^\prime)
\label{Dslash_F}
\end{eqnarray}
\begin{equation}
{\not \hspace{-3pt} D}^\bot(s,s^\prime) = \left\{ \begin{array}{ll} 
P_R \delta(1-s^\prime) - m_f P_L \delta(L_s-1 - s^\prime) - \delta(0 - s^\prime) & s=0 \\ 
P_R \delta(s+1 - s^\prime) + P_L \delta(s-1 - s^\prime) - \delta(s-s^\prime) & 0 < s < L_s-1 \\ 
-m_f P_R \delta(0-s^\prime) + P_L \delta(L_s-2 - s^\prime) - \delta(L_s-1 - s^\prime) & s = L_s -1
\end{array}
\right. 
\label{Dslash_perp_f}
\end{equation}
\begin{equation}
P_{R,L} = { 1 \pm \gamma_5 \over 2}
\end{equation}
where $V$ is the gauge field in the adjoint representation.
It is related to the field in the fundamental representation:
\begin{equation}
[V_\mu(x)]_{a,b} = \half {\rm Tr} [U^\dagger_\mu(x) \s^a U_\mu(x) \s^b].
\label{adj_fnd}
\end{equation}
with $\sigma^a$ the Pauli matrices.
In the above equations $m_0$ is a five-dimensional mass representing
the ``height'' of the domain wall and it controls the number of
light flavors in the theory.  In order to get one light species
in the free theory one must set $0<m_0<2$
\cite{Kaplan:1992bt}. The parameter $m_f$ explicitly mixes the two
chiralities and as a result it controls the bare fermion mass of the
four-dimensional effective theory.   In our simulations we
have set $m_f=0$.

The fermion field $\Psibar$ is not independent but is related to
$\Psi$ by the equivalent of the Majorana condition for this
5-dimensional theory \cite{Kaplan:1999jn}:
\begin{equation}
\Psibar = \Psi^T C R_5
\label{5D_Majorana_cond}
\end{equation}
where $R_5$ is a reflection operator along the fifth direction and 
$C$ the charge conjugation operator in Eucledean space which
can be set to:
\begin{equation}
C = \gamma_0 \gamma_2 \ .
\label{charge_conj}
\end{equation}
Therefore, the fermion action can also be written as:
\begin{equation}
S_F = - \sum_{x,x^\prime,s,s^\prime} \Psi^T(x,s) 
M_F(x,s; x^\prime,s^\prime) 
\Psi(x^\prime,s^\prime)
\label{action_F1}
\end{equation}
where
\begin{equation}
M_F(x,s; x^\prime,s^\prime) = C R_5 D_F(x,s; x^\prime,s^\prime) 
\label{antisym_frm_matrix}
\end{equation}
is an antisymmetric matrix as can be easily checked \cite{Neuberger:1997bg}.
As a result the fermionic integral gives the anticipated
Pfaffian:
\begin{equation}
\int [d\Psi] e^{-S_F} = {\rm Pf}(M_F)  \ .
\label{frm_pfaffian}
\end{equation}
Because $\det(C R_5) = 1$ one also has that $\det( M_F)  = \det( D_F ) $
and therefore:
\begin{equation}
{\rm Pf}(M_F) = \sqrt{\det( D_F )}  \ .
\label{frm_sqrt_det}
\end{equation}

The Pauli-Villars (PV) action $S_{PV}$ is designed to cancel the
contribution of the heavy fermions \cite{NN1}. Viewing the extra
dimension as an internal flavor space \cite{NN1} one can see that
there are $L_s-1$ heavy fermions with masses near the cutoff and one
light fermion. The PV subtraction removes effects of the $L_s$ heavy particles
in such a way that the overlap determinant is obtained in the
$L_s \to \infty$ limit \cite{Neuberger:1997bg}.  The PV subtraction
used here is as in \cite{PMV_Schwinger} and is given by:
\begin{equation}
S_{PV} =
\sum_{x,x^\prime,s,s^\prime} \Phi^T(x,s) 
M_F[m_f=1](x,s; x^\prime, s^\prime) \Phi(x^\prime,s^\prime)  \ .
\label{action_PV}
\end{equation}
The integral over the PV fields results in:
\begin{equation}
\int [d\Phi] e^{-S_{PV}} = {1 \over {\rm Pf}(M_F[m_f=1])}  \ .
\label{frm_pfaffian_pv}
\end{equation}

The gaugino condensate is measured using
four-dimensional fermion fields that are a projection
of the five-dimensional DWF \cite{Furman:1994ky}:
\begin{eqnarray} 
\chi(x)    &=& P_R \Psi(x,0) + P_L \Psi(x, L_s-1) \nonumber \\
\chibar(x) &=& \Psibar(x,L_s-1) P_R + \Psibar(x, 0) P_L  \ .
\label{projection}
\end{eqnarray} 
In the $L_s \to \infty$ limit of the theory these operators
directly correspond to insertions in the overlap of appropriate creation
and annihilation operators \cite{NN1}. 

Using Eq.~\myref{5D_Majorana_cond} and \myref{projection} the Majorana
condition on the four-dimensional fermion field is:
\begin{equation}
\chibar = \chi^T C  \ .
\label{4D_Majorana_cond}
\end{equation}
Because this is the correct condition for a four-dimensional field
one can see that the definition in Eq.~\myref{5D_Majorana_cond}
not only produces an antisymmetric fermion matrix $M_F$ but is
also consistent with the projection prescription in Eq.~\myref{projection}
as expected.

\section{Residual mass}
\label{smres}
Residual chiral symmetry breaking is understood
through the axial Ward identity on the DWF lattice \cite{Blum:2000kn},
which we write here for the case of vanishing bare mass:
\beq
\nabla_\mu \vev{ A_\mu^a(x) P^b(0)} = 2 \vev{J_{5q}^a(x) P^b(0)}
\eeq
where $A_\mu^a$ is a DWF version of the axial current with isospin
index $a=1,2,3$ and $P^a$ is a DWF version of the corresponding
pseudoscalar current, the interpolating operator for the ``pion.''
We note that in the present case the pion is in the adjoint
representation of the SU(2) gauge group (not to be confused with
isospin).  The pseudoscalar current $J_{5q}^a$ is different in
that it represents a pion at the middle of the fifth dimension:
\beq
J_{5q}^a(x) = -\psibar (x,L_s/2-1) P_L \s^a \psi(x,L_s/2)
+ \psibar(x,L_s/2) P_R \s^a \psi(x,L_s/2-1).
\eeq
It accounts for the difference in how the left- and right-handed
``quark'' fields in the DWF description transform, which causes
a mismatch midway between the domain walls and hence the explicit nonconservation
of the axial current in the lattice theory at finite $L_s$.

Since in the continuum limit the Ward identity must transition
to the continuum form, we see that $J_{5q}$ must be related to
the pseudoscalar current $P$ through the residual mass that is
a consequence of $L_s \not= \infty$:
\beq
J_{5q}^a \approx \mres P^a.
\eeq
Thus to extract the residual chiral symmetry breaking one studies
the large (imaginary) time behavior of the ratio:
\beq
\mres = \lim_{t\to\infty} \frac{ \sum_{\vec x, \vec y} J_{5q}^a(t,\vec x)
P^a(0,\vec y) }{ \sum_{\vec x, \vec y} P^a(t,\vec x)
P^a(0,\vec y) } .
\eeq
In all of our work we find that this quantity reaches a plateau
in the range $3 < t < T-2$, and fit to a constant in that region of $t$.

\section{Simulation}
\label{ssim}
Here we make a few brief remarks on the computational
aspects of this project.  Configurations were generated with the
rational hybrid Monte Carlo algorithm~\cite{rhmc98,rhmc03,rhmc04}.
All simulations were performed
on Rensselaer's Computational Center for Nanotechnology Innovations
cluster of 16 IBM BlueGene/L machines.  We typically used the full
capacity of two such machines 24 hours/day, and generated configurations for
approximately ten months, for a total of approximately
30 Million IBM BlueGene/L core hours.  The time required for data analysis was
a small fraction of this, by comparison.  Naturally, the large
lattices ($16^3 \times 32$) with small $\mres$ values were the
most costly to generate.

In the rational approximation used to generate configurations,
we found that it was necessary to go to rather high degrees,
due to a very wide spread between lowest and highest eigenvalues
of the Dirac operator.  This occured because we performed
our simulations at vanishing bare fermion mass $m_f=0$, relying on
the finite but large value of $L_s$ as an infrared regulator.
Typically, the Metropolis step required degrees between 15 and
20 in the computation of the change in the Hamiltonian, in order
to have sufficient accuracy.  Moreover, it was not unusual to
require between 50 and 100 steps in the leapfrog integration
for a trajectory of $\tau=0.5$ simulation time units, in order
to get reasonable acceptance rates at large $L_s$.  Again,
this was a result of small eigenvalues of the Dirac operator.
Naturally, these features led to very slow updating.  For the
$L_s$ values that we simulated, a single BlueGene/L rack was
able to produce $\ord{10}$ to $\ord{100}$ configurations per day.
Some examples are given in Table \ref{ttime}.  The last row
represents a run that was not reported in the main text as
it was too slow for a reasonable data set to be
generated in a practical time-frame.  As a result, our forthcoming work
will set $m_f \not =0$ and perform an $m_f\to 0$ extrapolation
for these larger volume, large $L_s$ simulations.

\begin{table}
\begin{center}
\begin{tabular}{|c|c|c|c|c|} \hline
$\beta$ & $V \times T$ & $L_s$ & $N_\tau$ & configs./day \\ \hline
2.3 & $8^3 \times 32$ & 48 & 60 & 140 \\
2.3 & $8^3 \times 32$ & 64 & 80 & 70 \\
2.3 & $16^3 \times 32$ & 32 & 90 & 40 \\
2.4 & $8^3 \times 32$ & 48 & 100 & 80 \\
2.4 & $16^3 \times 32$ & 48 & 200 & 11 \\ \hline
\end{tabular}
\caption{Example timing results on a single BlueGene/L rack (1024 dual
core nodes).  Here, $N_\tau$ is the number of steps in the
leapfrog trajectory. \label{ttime} }
\end{center}
\end{table}

\end{document}